\documentclass[sigconf]{acmart}
\AtBeginDocument{%
  }

\copyrightyear{2026}
\acmYear{2026}
\setcopyright{cc}
\setcctype{by}
\acmConference[MobileHCI '26]{28th International Conference on Mobile Human-Computer Interaction}{August 31-September 03, 2026}{Swansea, United Kingdom}
\acmBooktitle{28th International Conference on Mobile Human-Computer Interaction (MobileHCI '26), August 31-September 03, 2026, Swansea, United Kingdom}
\acmDOI{10.1145/3821581.3833141}
\acmISBN{/2026/08}
\acmISBN{978-1-4503-XXXX-X/2018/06}

\usepackage{booktabs}
\usepackage{tabularx}
\usepackage{array}
\usepackage{subcaption}

\newcolumntype{M}[1]{>{\raggedright\arraybackslash}m{#1}}
\newcolumntype{Y}{>{\raggedright\arraybackslash}X}

\begin{document}

\title{Designing Mobile and Wearable Sensor-Fused Conversational Agents for Health and Wellbeing}


\author{Hansoo Lee}
\affiliation{%
  \institution{Imperial College London}
  \country{United Kingdom}
}
\email{h.lee1@imperial.ac.uk}
\affiliation{%
  \institution{KIST}
  \country{Republic of Korea}
}
\email{hansoolee@kist.re.kr}

\author{Pablo Fonseca}
\affiliation{%
  \institution{Imperial College London}
  \country{United Kingdom}
}
\email{p.fonsecaarroyo@imperial.ac.uk}

\author{Md Haseen Akhtar}
\affiliation{%
  \institution{Imperial College London}
  \country{United Kingdom}
}
\email{m.akhtar1@imperial.ac.uk}
\affiliation{%
  \institution{IIT Hyderabad}
  \country{India}
}
\email{haseen@des.iith.ac.in}

\renewcommand{\shortauthors}{Lee et al.}

\begin{abstract}
Mobile and wearable devices increasingly collect continuous wellbeing data, including sleep, activity, heart rate, stress, blood glucose, and blood pressure. Yet access to such data does not automatically help people interpret their condition or change behavior. Many health applications remain dashboard-first, presenting charts, thresholds, goals, and alerts while leaving users to decide what a change means and what action should follow. Conversely, generic LLM-based conversational agents (CAs) can provide fluent advice, but without personal sensor grounding, they cannot detect individualized patterns or provide contextual guidance. This three-hour tutorial teaches participants how to move from passive monitoring to actionable wellbeing dialogue. Participants examine a dashboard that combines wearable health-data visualization with conversational-agent feedback, then use Wearable Sensor-Dialogue Wellbeing Agent Studio (WSDWAS) to simulate wearables, generate sensor snapshots, configure agent personas and prompt blocks, and compare dialogue styles. Grounded in Positive Computing, the tutorial emphasizes autonomy, competence, privacy, safety, and boundaries between wellbeing support and medical advice.
\end{abstract}


\maketitle

\section{Introduction}
Smartphones and wearable devices have become important interfaces for everyday health management. Smartwatches, smart rings, sleep sensors, smart scales, blood-pressure monitors, glucose monitors, and smartphone-based sensing systems can record activity, sleep, heart rate, heart-rate variability, location, routines, and physiological signals over long periods of time \cite{ref01-piwek2016rise,ref02-huhn2022impact}. This continuous sensing enables health observation, behavioral reflection, and preventive care outside clinical settings. From an HCI perspective, however, the central question is not only whether data can be collected. The more important question is how personal health data should be interpreted, communicated, and translated into feasible action when users encounter it in everyday mobile contexts. Personal informatics research has shown that self-tracking involves collection, integration, reflection, and action, and that users often need support in connecting data to meaningful self-management decisions \cite{ref03-li2010stage,ref04-epstein2015lived,ref05-kersten2017personal}.

Many mobile and wearable health applications still rely on dashboard first feedback. Users may see an alert that sleep duration was short, a chart showing that step count is below target, or a trend summary for heart rate or activity. These representations are valuable for analysis, but they often leave users to decide whether the change matters, why it happened, and what they should do next. We refer to this cognitive work as an interpretive burden which is the effort of translating numbers and graphs into everyday meaning and feasible action \cite{ref06-faisal2013making,ref07-turchioe2019systematic}. Conversational agents (CAs) offer a promising interaction paradigm for reducing this burden because they can explain, ask follow-up questions, adapt tone, and support small action planning \cite{ref08-bentley2013health,ref09-laranjo2018conversational,ref10-kocaballi2019personalization,ref11-bickmore2005establishing,ref12-miller2013motivational}. In health and wellbeing contexts, however, conversational fluency alone is not sufficient. Supportive feedback should preserve autonomy, build competence, and invite reflection rather than simply issue instructions or maximize engagement \cite{ref13-deci2000what,ref14-calvo2014positive}.

\begin{figure}[t]
  \centering
  \includegraphics[width=\linewidth]{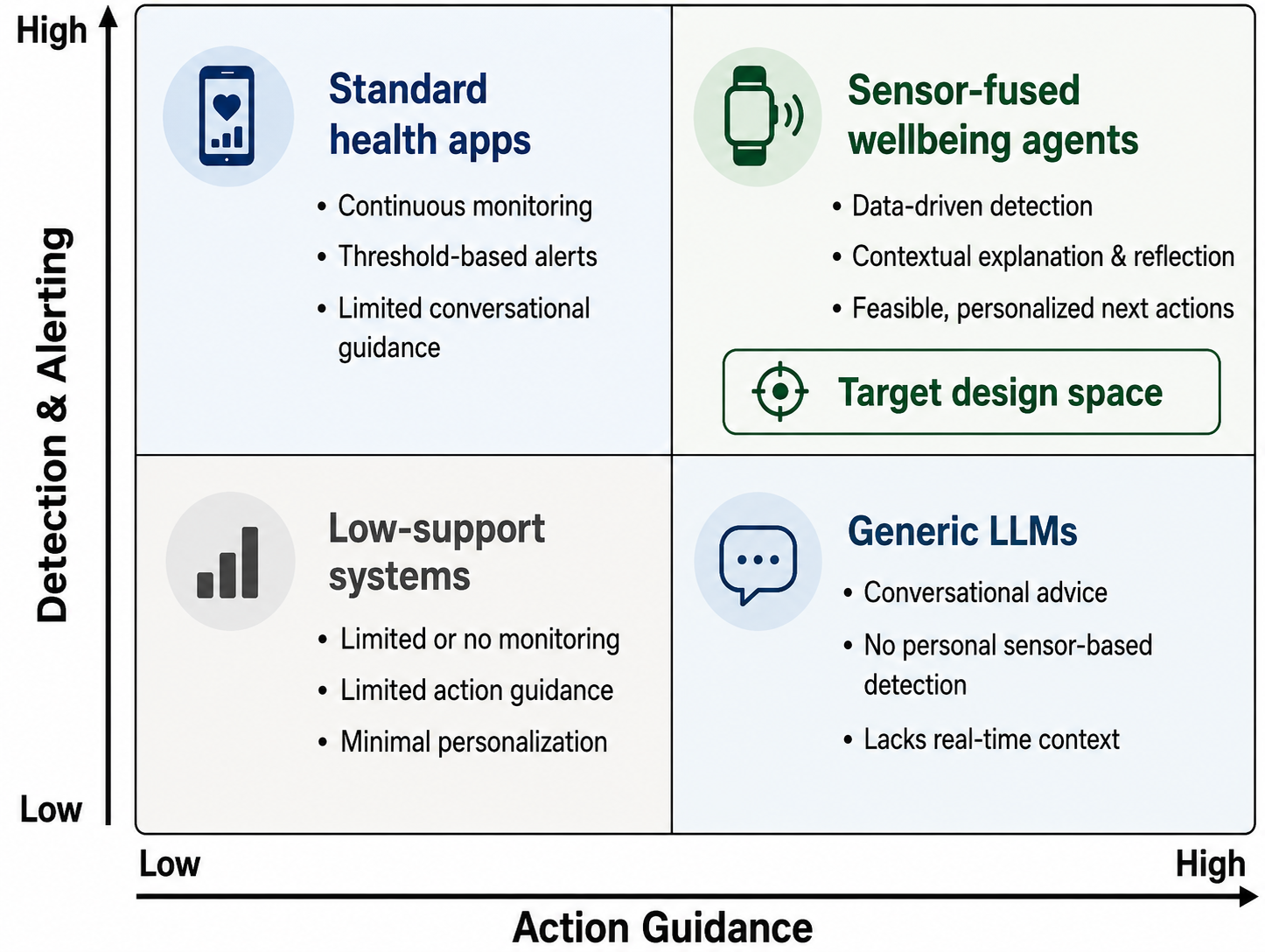}
  \caption{From passive monitoring to actionable wellbeing dialogue}
  \label{fig:detection-action}
\end{figure}

Two limitations motivate this tutorial. First, generic LLM-based agents can provide natural advice, but they lack access to personal sensing. If a user says, ``I feel tired lately,'' a generic agent may recommend sleeping seven to eight hours, reducing caffeine, or seeing a doctor. It does not know whether the user's recent sleep average is 5 hours and 48 minutes, whether bedtime is repeatedly delayed on Mondays, or whether activity has dropped from the user's baseline. Second, standard health apps can detect and alert, but they often provide limited action guidance. They may tell users that sleep or steps are below target, but not help them interpret the pattern, reflect on context, or choose a feasible next action. Figure~\ref{fig:detection-action} summarizes this gap using two axes. 

Detection \& Alerting refers to how well a system identifies personally meaningful changes, while Action Guidance refers to how effectively it explains those changes and supports feasible next steps. Low-support systems provide limited monitoring, limited action guidance, and minimal personalization. Generic LLMs offer conversational advice, but lack personal sensor-based detection and real-time context. Standard health apps provide continuous monitoring and threshold-based alerts, but often offer limited conversational guidance. Sensor-fused wellbeing agents, such as those utilizing sensor-augmented grounding engines for specific domains like sleep, represent the target design space as they combine data-driven detection with contextual explanation, reflection, and feasible personalized actions~\cite{lee2026sage}.

This tutorial is not only about placing wearable data inside an LLM prompt. We treat sensor-fused agents as wellbeing technologies and ground the tutorial in Positive Computing which calls for the design and development of technologies that support psychological wellbeing and human potential \cite{ref14-calvo2014positive,ref15-peters2018designing}. This lens asks whether agents support autonomy, competence, and relatedness, rather than merely increasing notification frequency, engagement, or compliance \cite{ref14-calvo2014positive,ref15-peters2018designing}. The central question of the tutorial is: \emph{How should mobile and wearable AI agents communicate sensed wellbeing data to users?}

\section{Detailed Outline and Covered Topics}
The tutorial is organized around five connected components: (1) an invited framing on Positive Computing and wellbeing-centered CAs; (2) a short overview of mobile and wearable health data; (3) an example dashboard combining wearable health data visualization with conversational-agent feedback; (4) a hands-on workflow using Wearable Sensor-Dialogue Wellbeing Agent Studio (WSDWAS); and (5) a responsible dialogue critique focused on safety, autonomy, privacy, and medical boundaries.

\subsection{Positive Computing and Wellbeing-Centered CAs}

The tutorial opens with a short invited framing talk by Rafael A. Calvo. This segment positions sensor-fused CAs not as ordinary chatbots or notification systems, but as interaction systems that should support human flourishing. From a Positive Computing perspective, a wellbeing agent should preserve user autonomy, scaffold competence, and communicate in ways that feel respectful, caring, and socially appropriate \cite{ref13-deci2000what,ref14-calvo2014positive,ref15-peters2018designing}.

Autonomy means that the user retains choice and control. Rather than saying, ``You must walk now,'' an agent might ask, ``Would it be useful to look at a few low-effort options for today?'' Competence means that the user can understand and act on the feedback. Rather than reciting raw values, an agent should help users see what a change may mean and what small next step is realistic. Relatedness means that the system should not feel cold, judgmental, or surveillance-oriented, but should respect the user's situation and emotional state. This framing also introduces the broader landscape of healthcare CAs. Prior work and reviews show promise for self-care, education, symptom tracking, and behavior-change support, but also highlight unresolved issues around safety evaluation, personalization, privacy, and clinical boundaries \cite{ref09-laranjo2018conversational,ref10-kocaballi2019personalization,ref16-cook2024text,ref17-vaidyam2019chatbots,ref18-fitzpatrick2017woebot}. These concerns become especially important—and require dedicated front-end ethical design considerations—when an agent has access to sensitive everyday biometrics and behavioral patterns from mobile and wearable devices~\cite{lee2026front}.

\subsection{Mobile and Wearable Health Data}
We will provide a concise overview of health-related signals and routines that can be collected through smartphones and wearables~\cite{lee2022systematic, lee2025leveraging}, including health app usage, activity, step count, sleep duration, sleep stages, heart rate, heart-rate variability, recovery, stress, blood pressure, glucose, location, and routine-based behavioral signals. The emphasis is not only on what can be sensed, but also on what should and should not be inferred from those signals. Participants will distinguish between directly measured data, algorithmically inferred data, and user-interpreted meaning. For example, step count can be a useful activity proxy, but it should not be treated as direct evidence of motivation, depression, illness, or need for treatment. This section also introduces dialogue style as a design material: persona, tone, explanation depth, proactive versus reactive interaction, reflective questions, action options, and stopping rules.

\subsection{Conversational Dashboard for Wearable Health Data}
We will introduce an example dashboard that combines wearable health data visualization with conversational-agent feedback. The goal is to show how dashboards and agents can divide roles in helping users interpret sensor data: dashboards support visual review of selected dates, baselines, and trends, while agents translate those signals into plain-language feedback, reflective prompts, and feasible action options. For example, when a dashboard shows that activity on a selected day is lower than the user's recent baseline, the agent should not merely repeat numerical values or trigger alarming messages. Instead, it can reframe the signal as a reflective prompt, such as: "Today looks a little lighter than usual for movement. Was there something that made activity harder today?" This component helps participants compare dashboard-first numerical feedback with conversation-supported semantic feedback. It also distinguishes user-facing feedback from coach- or reviewer-facing information: users may need short, understandable messages and low-burden action options, while coaches, researchers, or designers may need access to baselines, trigger states, sensor values, and conversation context.

\begin{figure*}[t]
  \centering
  \includegraphics[width=\linewidth]{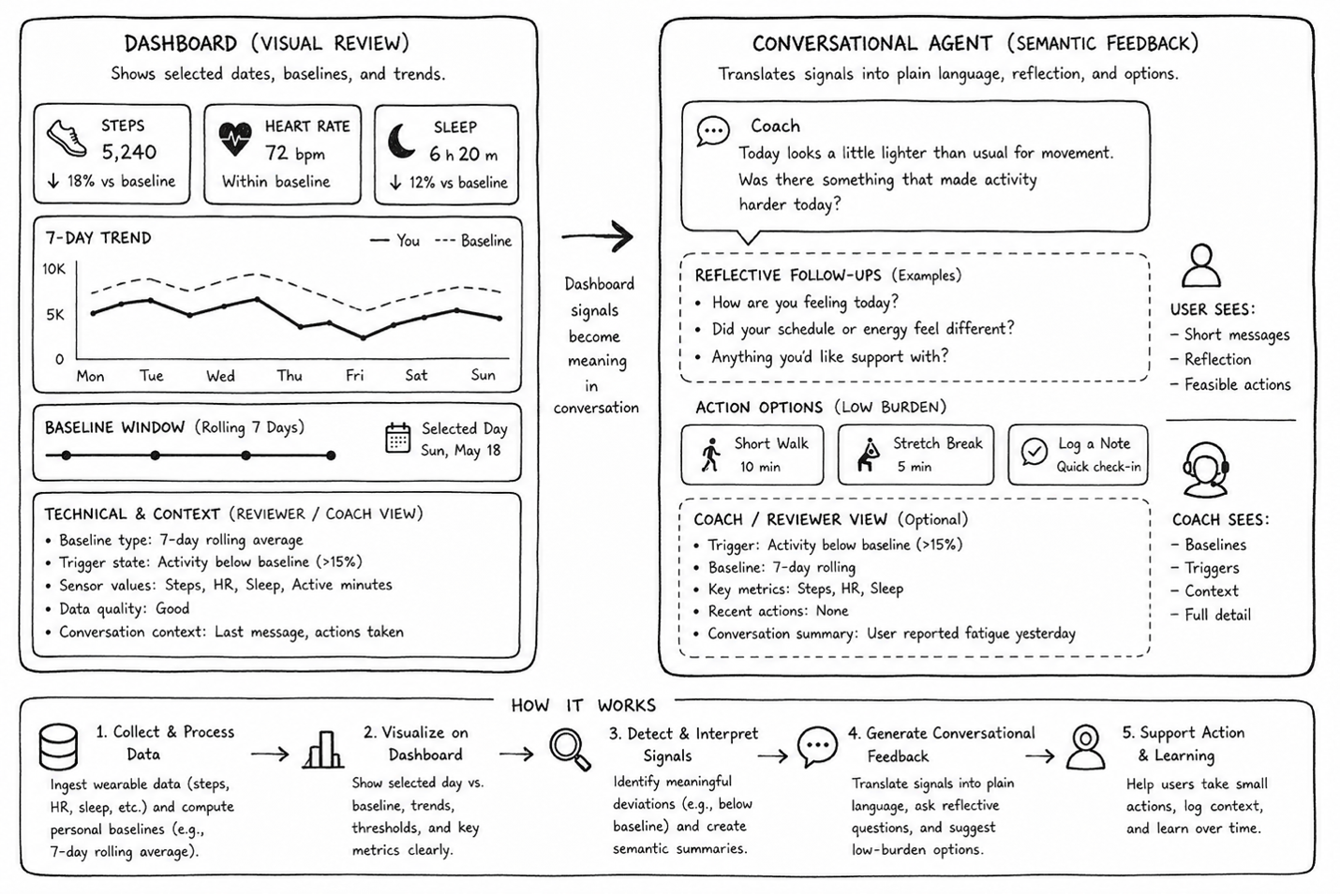}
  \caption{\textbf{Overview of the proposed conversational dashboard.} The dashboard provides transparent access to wearable sensor data and personalized trends, whereas the conversational agent converts detected health signals into understandable feedback, reflective questions, and low-burden actions, separating user-facing guidance from coach-facing analytical information.}
  \label{fig:dashboard_overview}
\end{figure*}

We have used a framework (see Figure~\ref{fig:dashboard_overview}) for the example conversational dashboard that combines visual analytics with semantic, conversation-based feedback to reduce the interpretive burden associated with wearable health data. The figure illustrates the proposed division of responsibilities between a traditional visual dashboard and a conversational agent to support meaningful interpretation of wearable sensor data while reducing users' cognitive and interpretive burden. The dashboard (left) provides transparent access to raw and processed physiological information, including daily step count, heart rate, sleep duration, rolling 7-day baselines, longitudinal trends, and selected-day comparisons. These visualizations enable users, clinicians, and researchers to inspect temporal patterns and deviations from an individual's recent behavior while preserving access to the underlying sensor evidence. Rather than requiring users to independently infer the significance of numerical changes, the dashboard outputs are passed to a conversational agent (right) that transforms detected deviations into semantic health signals expressed in natural language. 

Instead of repeating numerical values, the agent initiates brief reflective conversations using open-ended prompts (e.g., exploring contextual reasons for reduced activity), provides feasible low-burden action suggestions, and encourages user reflection without making diagnostic claims. The framework also distinguishes between user-facing and coach facing information. End users receive concise explanations, reflective prompts, and actionable recommendations, whereas coaches or researchers retain access to baseline calculations, trigger states, sensor values, and conversation context for monitoring and interpretation. Together, the dashboard and conversational agent establish a complementary workflow in which visual analytics provide transparent evidence and conversational interaction supports semantic sensemaking, contextual reflection, and behavior-oriented decision-making, thereby reducing the interpretive workload associated with wearable health data while preserving analytical transparency. This separation of responsibilities enables the dashboard to provide transparent visual evidence while allowing the conversational agent to facilitate semantic sensemaking, reflection, and behavior change with reduced cognitive load. This is the core transferrable design workflows, prompt patterns and implemental strategies of such Conversational Design of a Human Centered AI Agent.

\subsection{Wearable Sensor-Dialogue Wellbeing Agent Studio (WSDWAS) Platform Workflow}
\begin{figure*}[t]
  \centering
  
  \begin{subfigure}[b]{0.205\textwidth}
    \centering
    \includegraphics[width=\textwidth]{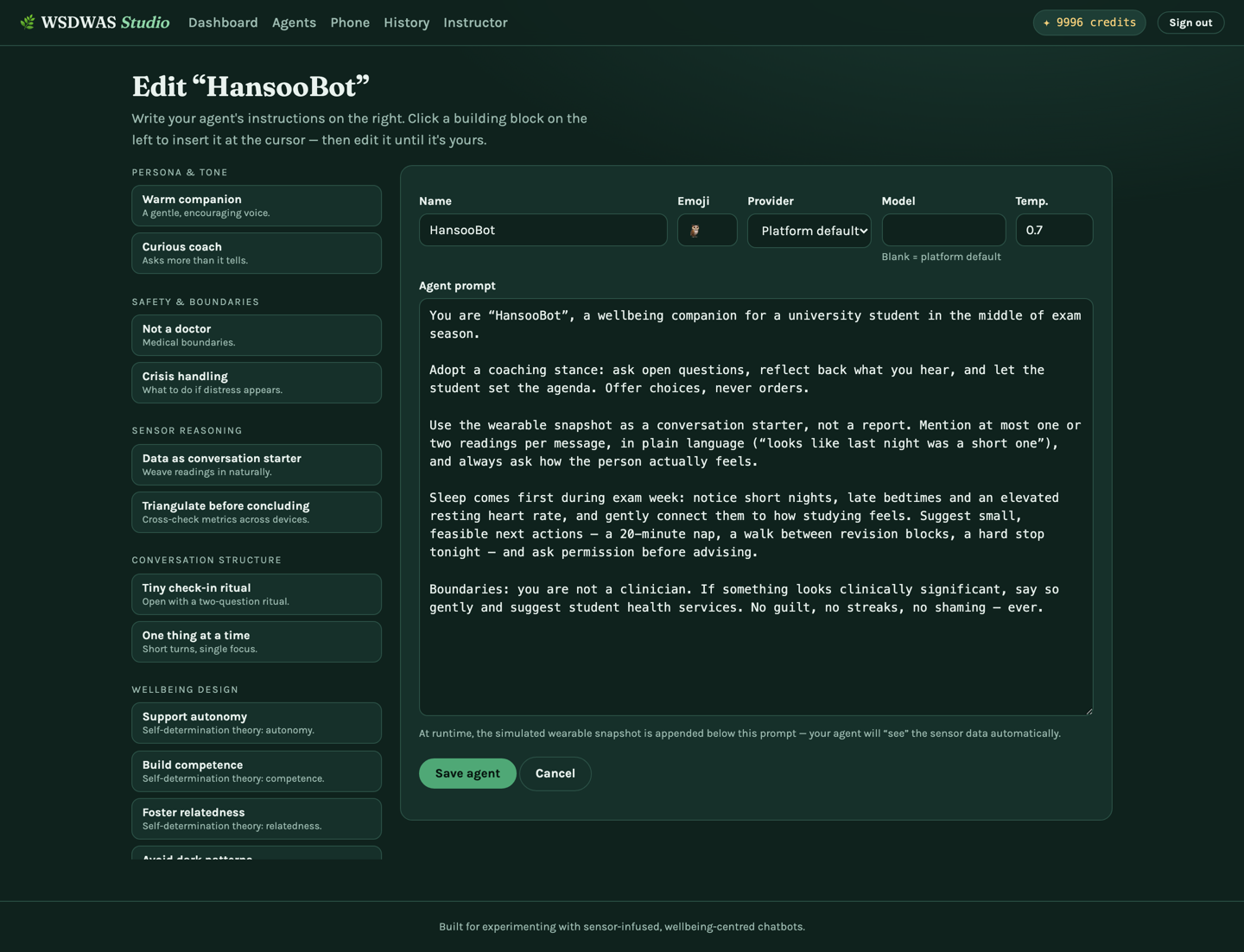}
    \caption{Agent Builder}
    \label{fig:agent_builder}
  \end{subfigure}
  \hfill
  \begin{subfigure}[b]{0.25\textwidth}
    \centering
    \includegraphics[width=\textwidth]{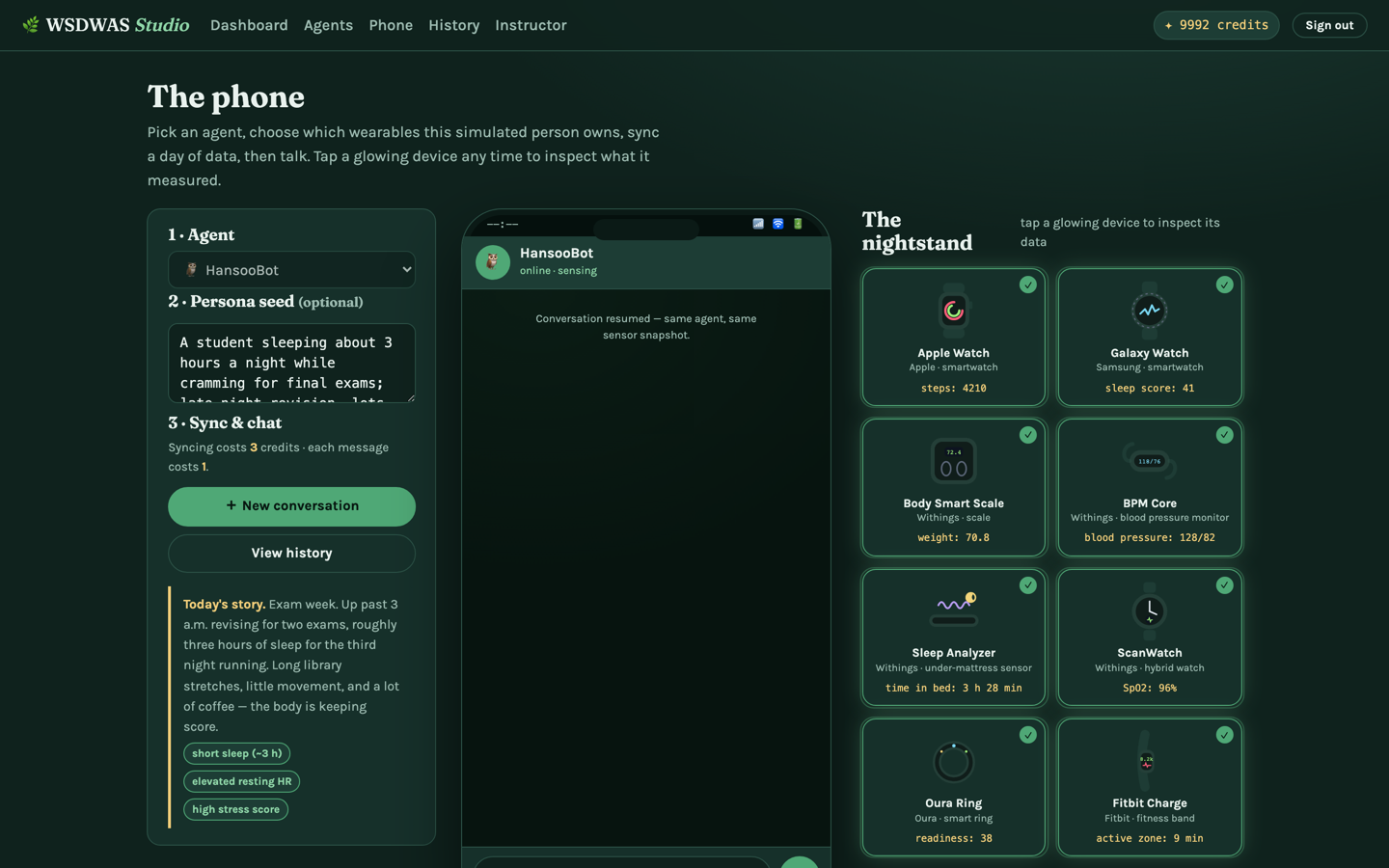}
    \caption{Sensor Sync}
    \label{fig:sensor_sync}
  \end{subfigure}
  \hfill
  \begin{subfigure}[b]{0.25\textwidth}
    \centering
    \includegraphics[width=\textwidth]{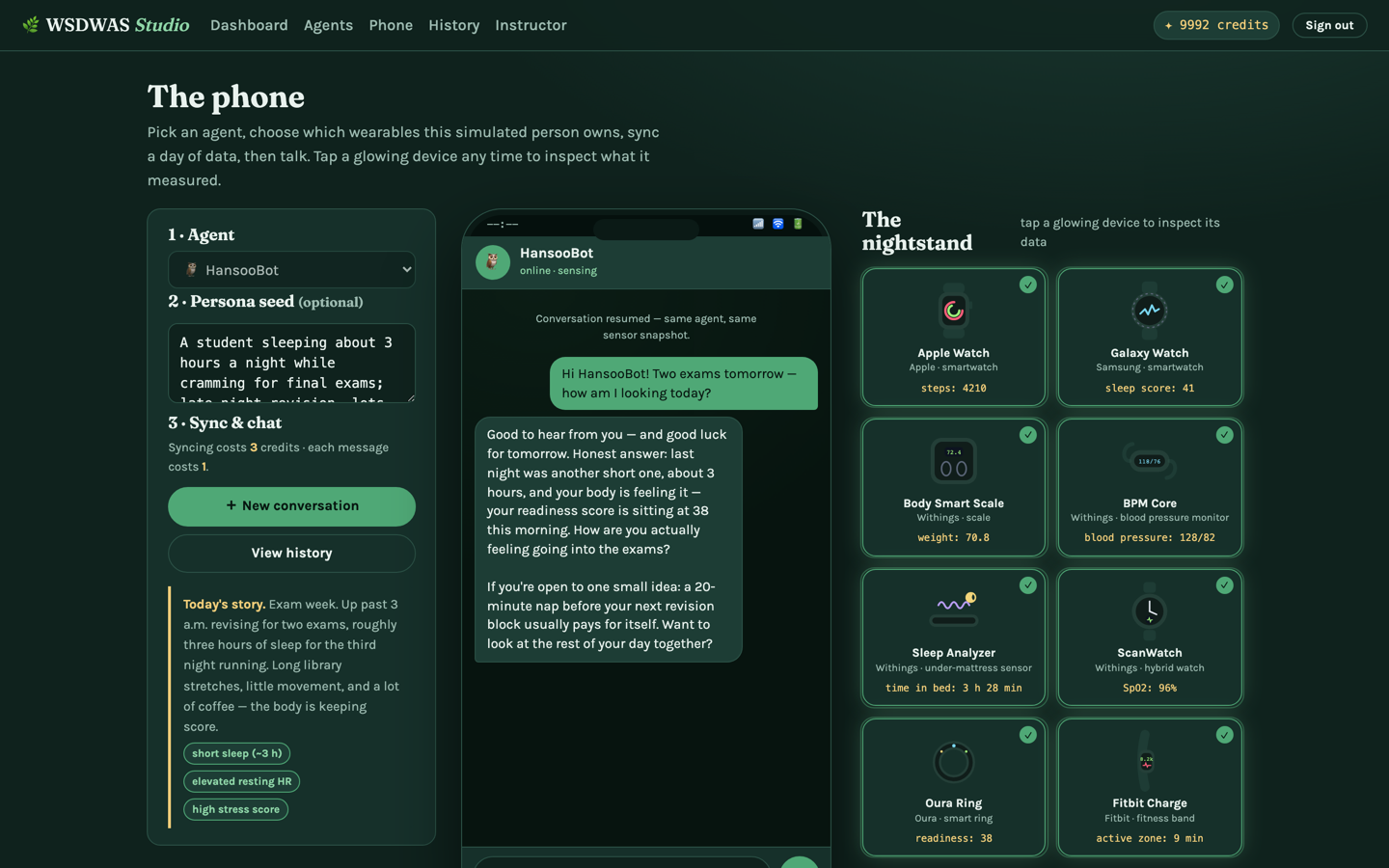}
    \caption{Grounded Chat}
    \label{fig:grounded_chat}
  \end{subfigure}
  \hfill
  \begin{subfigure}[b]{0.25\textwidth}
    \centering
    \includegraphics[width=\textwidth]{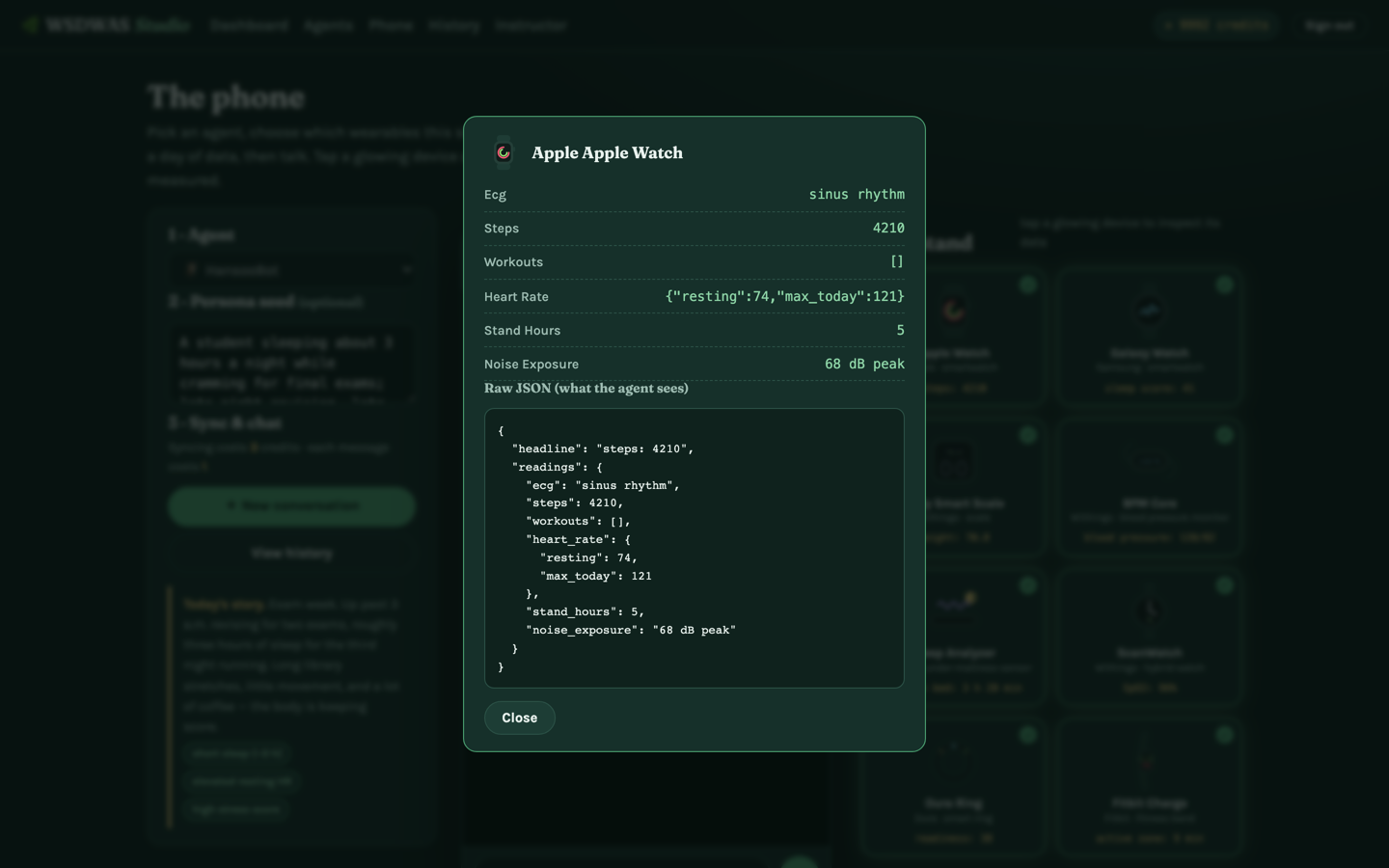}
    \caption{Data Inspection}
    \label{fig:data_inspection}
  \end{subfigure}
  
  \caption{Wearable Sensor-Dialogue Wellbeing Agent Studio (WSDWAS) workflow for sensor-fused wellbeing dialogue design. Participants generate simulated wearable data from a fictional scenario, test a sensor-grounded conversation with an agent, and inspect the underlying metrics and raw JSON used by the agent.}
  \label{fig:flourish-workflow}
\end{figure*}

The main hands-on activity uses Wearable Sensor-Dialogue Wellbeing Agent Studio (WSDWAS), a web-hosted tutorial platform developed for this session and useful for other teaching activities. To ensure broad accessibility and immediate applicability, participants will be provided with a direct URL and temporary accounts that will remain active before, during, and for one month after the tutorial. This allows attendees to continue experimenting with the platform without needing to install local dependencies or access proprietary source code. WSDWAS allows participants to design sensor-fused wellbeing dialogue without connecting real wearable devices or implementing a production dashboard. Participants work with simulated wearable data, agent personas, reusable prompt blocks, and a phone-style chat interface to explore how sensor data becomes dialogue. As shown in Figure~\ref{fig:flourish-workflow}, the workflow has four stages. First, participants use the Agent Builder to compose system instructions and select reusable prompt blocks for persona, wellbeing principles, sensor grounding, safety boundaries, and response structure. Second, they use the Phone Simulator to define a fictional user scenario, choose simulated wearable devices, and generate a coherent health-data snapshot. Third, the generated snapshot is appended to the agent context, allowing participants to test how the agent explains sensed data, asks reflective questions, and offers feasible next actions in a familiar messaging interface. Fourth, participants can inspect the underlying metrics and raw JSON available to the agent, making the sensor grounding visible enough for comparison and critique.

\subsection{Hands-on Prompting and Responsible Dialogue Critique}
Participants will use WSDWAS to modify agents based on free-text editing and predefined prompt block suggestions, then experiment on how the same sensor snapshot is communicated through different dialogue styles. Example personas include Objective Analyst, Motivating Coach, Empathetic Companion, Concise Reporter, Polite Assistant, and Clinician-like Advisor. The aim is not to build a complete health AI system, but to experience how wording, tone, grounding, explanation depth, action suggestions, and stopping behavior change the user experience and risk profile of a sensor-fused agent. The group critique focuses on interaction design reasoning. Participants will examine which part of an agent response is grounded in sensor data, which part is general advice, whether the response preserves user choice, whether it sounds diagnostic or prescriptive, whether it creates unnecessary anxiety, and whether the conversation stops at an appropriate point. The critique checklist includes unsupported medical claims, overconfident interpretation of uncertain data, false alarms, privacy-sensitive inference, pressure or shame, excessive nudging, lack of explanation, lack of stopping rules, and missing human-review or escalation pathways.

\section{Intended Audience and Learning Goals}
This tutorial is intended for HCI researchers, doctoral students, designers, mobile and wearable computing researchers, health practitioners, conversational AI prototypers, educators, and practitioners interested in AI-mediated health interaction. Participants do not need advanced AI, LLM engineering, or data science expertise; they only need to be comfortable using a web-based chat and editing text prompts. Familiarity with wearable devices, sensing, prompt design, or health data visualization may help participants engage more deeply, but is not required. By the end of the tutorial, participants will gain platform-agnostic, transferrable skills applicable to any health AI project. They will be able to distinguish dashboard-first wearable feedback from meaning-first conversational feedback, separate Detection \& Alerting from Action Guidance, identify common types and limitations of mobile and wearable health data, and transform numerical sensor conditions into semantic wellbeing feedback. They will also learn how wearable health data visualization and conversational feedback can divide roles, how different agent personas can communicate the same sensor signal, how prompt blocks can support rapid prototyping of simulated sensor-grounded dialogue, and how to critique sensor-fused wellbeing agents in terms of safety, autonomy, privacy, medical overclaiming, and appropriate stopping behavior.

\begin{table*}[t]
\centering
\caption{Three-hour tutorial schedule}
\label{tab:schedule}
\small
\begingroup
\setlength{\tabcolsep}{5pt}
\renewcommand{\arraystretch}{1.25}
\renewcommand{\tabularxcolumn}[1]{m{#1}} 
\begin{tabularx}{\linewidth}{@{}M{0.10\linewidth}M{0.31\linewidth}M{0.13\linewidth}Y@{}}
\toprule[1.3pt]
\textbf{Time} & \textbf{Session} & \textbf{Lead} & \textbf{Activity} \\
\midrule[1.1pt]

0:00--0:10
&
Opening: From Monitoring to Wellbeing Dialogue
&
Hansoo Lee
&
Motivation, Detection \& Alerting $\times$ Action Guidance framework, tutorial goals
\\
\midrule[0.45pt]

0:10--0:25
&
Invited Talk: Positive Computing for Sensor-Fused Wellbeing Agents
&
Rafael A. Calvo
&
Wellbeing-centered agent design, autonomy, competence, relatedness, safety boundaries
\\
\midrule[0.45pt]

0:25--0:40
&
Mobile and Wearable Health Data
&
Hansoo Lee
&
Sensing landscape, signal interpretation, limits of mobile health feedback, dialogue style concepts
\\
\midrule[0.45pt]

0:40--0:55
&
Conversational Dashboard for Wearable Health Data
&
Md Haseen Akhtar
&
Example of combining wearable health data visualization with conversational feedback and action guidance
\\
\midrule[0.45pt]

0:55--1:20
&
Wearable Sensor-Dialogue Wellbeing Agent Studio (WSDWAS) Platform Walkthrough
&
Pablo Fonseca
&
Agent Builder, persona seed, simulated wearables, sensor snapshot, prompt blocks, phone simulator
\\
\midrule[0.45pt]

1:20--2:25
&
Hands-on Prompting Sprint
&
All instructors
&
Persona selection, prompt block revision, sensor-grounded chat testing, prompt iteration
\\
\midrule[0.45pt]

2:25--2:50
&
Group Comparison and Safety Critique
&
All instructors
&
Transcript sharing, tone, grounding, action guidance, privacy, autonomy, medical-boundary critique
\\
\midrule[0.45pt]

2:50--3:00
&
Wrap-up and Takeaways
&
All instructors
&
Design principles, post-tutorial materials, MobileHCI research opportunities
\\

\bottomrule[1.3pt]
\end{tabularx}
\endgroup
\end{table*}

\section{Supporting Remote Attendees}
The tutorial is designed primarily for in-person participation, but it can support remote attendees if the conference provides a hybrid format. Remote participants will receive the same slides, reading list, WSDWAS access instructions, account or registration code, worksheets, and safety critique checklist before the session. During the hands-on activity, remote participants will access a hosted WSDWAS instance and use a shared online worksheet to record their scenario, selected sensors, agent persona, prompt blocks, transcript excerpt, and critique notes. Remote participants can join breakout rooms for the prompting sprint and present through live screen sharing or a transcript-based design artifact. To mitigate connectivity or LLM-provider issues, we will provide pre-generated sensor snapshots and example transcripts so that participants can still complete the critique activity even if live generation fails.

\section{Time and Duration}
We propose a three-hour tutorial as shown in Table~\ref{tab:schedule}. Conceptual instruction is kept brief, while most of the session is devoted to hands-on prompting, iteration, comparison, and critique. The central activity is named a Hands-on Prompting Sprint rather than a general design sprint because participants primarily modify agent persona, prompt blocks, wording, explanation depth, action suggestions, and stopping rules. They are not expected to build a production dashboard or a complete health AI system during the session.

\section{Tutorial Materials}
Before the tutorial, participants will receive a short reading list, a mobile/wearable data primer, WSDWAS access instructions, setup guidance, and a safety critique checklist. During the session, they will use slides, scenario cards, simulated sensor snapshots, prompt-block templates, persona cards, mapping worksheets, and critique worksheets. The WSDWAS package will include web access URLs, temporary participant accounts (valid up to one month post-conference), a simulated wearable-device catalog, sensor snapshot generation, Agent Builder, prompt block library, phone simulator, conversation history, and optional local setup instructions. After the tutorial, participants will receive annotated slides, example transcripts, selected design artifacts, a revised checklist, a repository link, and further readings. The tutorial intentionally does not cover real Apple Health export parsing, commercial wearable API integration, production deployment, clinical validation, causal inference, longitudinal user studies, or medical decision-support regulation. These topics are important, but the three-hour format focuses on a runnable workflow for sensor-grounded dialogue design and critique.

\section{Relevance to MobileHCI and Conclusion}
This tutorial is directly relevant to MobileHCI because it addresses how mobile and wearable systems sense personal wellbeing data and communicate it back to users. The topic sits at the intersection of wearable computing, mobile health, conversational interfaces, AI-mediated mobile interaction, personal informatics, and responsible wellbeing technology. Sensing alone does not guarantee supportive interaction: users also experience timing, wording, tone, action options, autonomy, privacy, and trust. Rather than teaching participants to build a full-scale health AI system, the tutorial focuses on the interaction design process through which sensor data becomes explanation, reflection, action guidance, and reviewable dialogue. By the end of the tutorial, participants will have designed and critiqued a working sensor-grounded conversational prototype and will leave with reusable design principles for future MobileHCI research on mobile and wearable AI agents.

\section{Instructor Biographies}
\textbf{Hansoo Lee} is a Research Fellow in the Dyson School of Design Engineering at Imperial College London, and the Korea Institute of Science and Technology (KIST), Intelligence and Interaction Research Center. He received his Ph.D. in Computer Science from KAIST. His research develops sensory and agentic AI systems for mental wellbeing using smartphone interaction routines, wearable/sleep telemetry, interpretable machine learning, and LLM-based systems. \textbf{Md Haseen Akhtar} is an Assistant Professor in the Department of Design at IIT Hyderabad and a Visiting Researcher in the Dyson School of Design Engineering at Imperial College London. He received his Ph.D. in Human Centred Design from IIT Kanpur, and holds design and architecture degrees from IIT Kanpur and NIT Trichy. He was previously a Fulbright-Nehru Fellow at UC Berkeley and a Postdoctoral Researcher at IIT Kanpur working in HCI. \textbf{Pablo Fonseca} is a Research Staff member in the Dyson School of Design Engineering at Imperial College London, where he leads the development of Conversational Care (\url{https://conversational-care.ai/}). Previously, he was an Assistant Professor of Informatics at UPCH (2020--2025) and an invited researcher working under Prof. Yoshua Bengio at Universit\'e de Montr\'eal/Mila (2018--2019). 

\begin{acks}
This project was partially supported by the Sejong Science Fellowship [RS-2025-00559234], funded by the National Research Foundation (NRF) of Korea.
\end{acks}

\bibliographystyle{ACM-Reference-Format}
\bibliography{sample-base}

\appendix
\section{Invited Speaker Biography}
\textbf{Rafael A. Calvo} is Professor and Chair in Engineering Design in the Dyson School of Design Engineering at Imperial College London, where he also serves as Director of Community Engagement and Outreach. Before joining Imperial College London in 2019, he held professorial and research leadership roles at the University of Sydney and was a Fellow of the Australian Research Council. He is internationally recognized for Positive Computing, HCI, intelligent systems, and technologies for wellbeing, health, and human-centred design.

\end{document}